\documentstyle[12pt,amstex,amssymb]{article}
\input{feynman}
\def\alf1{ {\alpha\over\pi} }

\begin{document}

\begin{titlepage}
 \begin{flushright}
 
 {\bf UTHEP-96-0702 }\\
 {\bf July 1996}\\
\end{flushright}
\vspace{0.5cm}
 
\begin{center}
{\LARGE
Gauge Invariant YFS Exponentiation of (Un)stable \\
$Z$-Pair Production At and Beyond LEP2 Energies$^{\dagger}$
}
\end{center}

\begin{center}
 {\bf S. Jadach}\\
   {\em Institute of Nuclear Physics,
        ul. Kawiory 26a, Krak\'ow, Poland}\\
   {\em CERN, Theory Division, CH-1211 Geneva 23, Switzerland,}\\
 {\bf W.  P\l{a}czek}$^{\star}$\\
   {\em Department of Physics and Astronomy,\\
   The University of Tennessee, Knoxville, Tennessee 37996-1200},\\
 {\bf B.F.L. Ward}\\
   {\em Department of Physics and Astronomy,\\
   The University of Tennessee, Knoxville, Tennessee 37996-1200,\\
   SLAC, Stanford University, Stanford, California 94309}
\end{center}

\vspace{0.25cm}
\begin{center}
{\bf   Abstract}
\end{center}
\baselineskip=25pt 
We present the theoretical basis and sample Monte Carlo data for the YFS
exponentiated calculation of $e^+e^- \rightarrow
Z Z \rightarrow  f_1\bar f_1 + \bar f_2 f_2$ at and beyond
LEP2 energies, where the left-handed part of
$f_i$ is a
component of an $SU_{2L}$ doublet, $i=1,2$.
The calculation is performed for both SM couplings and for
anomalous $ZZV$ triple gauge boson couplings in the conventions
of Hagiwara {\it et al.}. Our formulas, which are gauge invariant,
are illustrated in a proto-typical YFS Monte Carlo
event generator YFSZZ.
\vspace{0.25cm}
\begin{center}
{\it To be submitted to Phys. Rev. D}
\end{center}
 
\vspace{0.5cm}
\renewcommand{\baselinestretch}{0.1}
\footnoterule
\noindent
{\footnotesize
\begin{itemize}
\item[${\dagger}$]
Work partly supported by the Polish Government
grant KBN 2P30225206
by the US Department of Energy Contracts  DE-FG05-91ER40627
and   DE-AC03-76ER00515.

\item[${}^{\star}$]
On leave of absence from Institute of Computer Science, Jagiellonian
University, Krak\'ow, Poland.
\end{itemize}
}
 
\begin{flushleft} 
{\bf UTHEP-96-0702 }\\
{\bf July 1996}\\
\end{flushleft}

\end{titlepage}
\baselineskip=25pt 
 

The problem of the precision calculation of the
processes $e^+e^- \to Z Z +n(\gamma)\to 4 fermions+n(\gamma)$
at the higher energy range of LEP2 and at NLC ($>300$ GeV) type
energies is the subject of our paper.
We are motivated to investigate this problem for a number of reasons,
the most important of which is its role
in connection with the
verification and tests of the $SU_{2L}\times U_1$ model of Glashow,
Salam and Weinberg~\cite{gsw} of the electroweak interaction.
Indeed, at the NLC these and related processes are expected to play
the primary role in such physics studies~\cite{nlc,bark} 
as precision
tests of the fundamental non-Abelian triple and quartet gauge field
self-interactions in principle, for example. In this paper, we present
and illustrate, via sample Monte Carlo data,
the rigorous Yennie-Frautschi-Suura (YFS)~\cite{yfs} exponentiated
Monte Carlo approach~\cite{sjw} to
these processes. The respective Monte Carlo event generator which realizes the
calculation is
exact in the infrared regime and is of leading logarithmic accuracy
through ${\cal O}(\alpha^2)$ in the initial state 
QED hard radiative regime. A higher
precision realization of our methods and results is in progress~\cite{elsewh}.

We recall at this time the current state of the art insofar as
Monte Carlo event generators for the processes 
$e^+e^- \to Z Z +n(\gamma)\to 4 fermions+n(\gamma)$ are concerned.
In ref.~\cite{lep2wkgrp,ohl}, both semi-analytical 
and Monte Carlo event generator results in the 
literature have been reviewed. What is new
in our work in this paper is that we present
for the first time the results of a calculation 
of theese processes in which all of the
following are simultaneously featured:
\begin{itemize}
\item realistic, finite $p_T$, simulation of the
respective 
multiple photon radiative effects on an event-by-event basis
in which infrared singularities are cancelled to all
orders in~$\alpha$
\item YFS exponentiated ${\cal O}(\alpha^2)$ LL initial state radiation (ISR)
\item anomalous triple gauge boson couplings.
\end{itemize}
Thus, our work represents the first realistic multi-photon
radiative effects simulation of the processes under study
in which the interplay between the $n(\gamma)$ radiation and
possible anomalous coupling effects can be systematically investigated.
We illustrate this point in what follows.

Our work is organized as follows. We first 
build on the recent application of the YFS exponentiated
Monte Carlo algorithm in ref.~\cite{yfs2} to the $W$-pair production 
process by
extending it to the $Z$-pair production process of
interest to us here. We then present
some sample MC data and discuss their implications for $Z$-pair
production studies at high energy $e^+e^-$ colliding beam
devices. Finally, we present some summary remarks.\par

\begin{figure}[!ht]
\centering
\begin{picture}(48000,27000)
\thicklines
\bigphotons
\put(20000,12000){\oval(4000,9000)}
\put(18500,16000){\line(1,0){3000}}
\put(18200,15500){\line(1,0){3600}}
\multiput(18000,15000)(0,-500){13}{\line(1,0){4000}}
\put(18500, 8000){\line(1,0){3000}}
\put(18200, 8500){\line(1,0){3600}}
\THICKLINES
\drawline\fermion[\SE\REG](13500,20500)[6700]
\drawarrow[\NW\ATBASE](15000,19000)
\drawline\photon[\NE\FLIPPEDFLAT](\pmidx,\pmidy)[7]
\drawarrow[\E\ATBASE](18200,20500)
\put(11500,20000){\large $e^+$}
\put(13000,17500){\large $-p_1$}
\put(20500,22000){\large $\gamma_1$}
\put(17000,21500){\large $k_1$}
\drawline\fermion[\NE\REG](13500, 3500)[6700]
\drawarrow[\NE\ATTIP](15000, 5000)
\drawline\photon[\SE\FLAT](\pmidx,\pmidy)[7]
\drawarrow[\E\ATBASE](18200,3500)
\put(11500, 3000){\large $e^-$}
\put(14000, 6000){\large $q_1$}
\put(20500, 1500){\large $\gamma_n$}
\put(17000, 2000){\large $k_n$}
\drawline\photon[\E\REG](21500,16000)[7]
\drawline\fermion[\NE\REG](\photonbackx,\photonbacky)[3000]
\drawarrow[\NE\ATBASE](\pmidx,\pmidy)
\drawline\fermion[\SE\REG](\photonbackx,\photonbacky)[3000]
\drawarrow[\NW\ATBASE](\pmidx,\pmidy)
\put(26000,17000){\large $Z$}
\put(26000,14500){\large $p_2$}
\put(31000,17500){$f_1$}
\put(31000,14000){$\bar{f_1}$}
\put(28500,18500){ $p_{f_1}$}
\put(27500,13500){ $-p_{\bar{f_1}}$}
\drawline\photon[\E\FLIPPED](21500, 8000)[7]
\drawline\fermion[\NE\REG](\photonbackx,\photonbacky)[3000]
\drawarrow[\SW\ATBASE](\pmidx,\pmidy)
\drawline\fermion[\SE\REG](\photonbackx,\photonbacky)[3000]
\drawarrow[\SE\ATBASE](\pmidx,\pmidy)
\put(26000, 9000){\large $Z$}
\put(26000, 6500){\large $q_2$}
\put(31000, 9500){$\bar{f_2}$}
\put(31000, 6000){$f_2$}
\put(27500,10500){ $-p_{\bar{f_2}}$}
\put(28500, 5500){ $p_{f_2}$}
\multiput(20000,20500)(1000,-1000){5}{\circle*{350}}
\multiput(24200, 9000)(   0, 1500){5}{\circle*{350}}
\multiput(20000, 3500)(1000, 1000){5}{\circle*{350}}

\end{picture}
 
\caption{\small\sf
The process
\protect$e^+e^-\rightarrow ZZ + n(\gamma)
\rightarrow  f_1+\bar{f_1} +\bar{f_2} + f_2 + n(\gamma)$. 
}
\label{fig:ZZprod}
\end{figure}
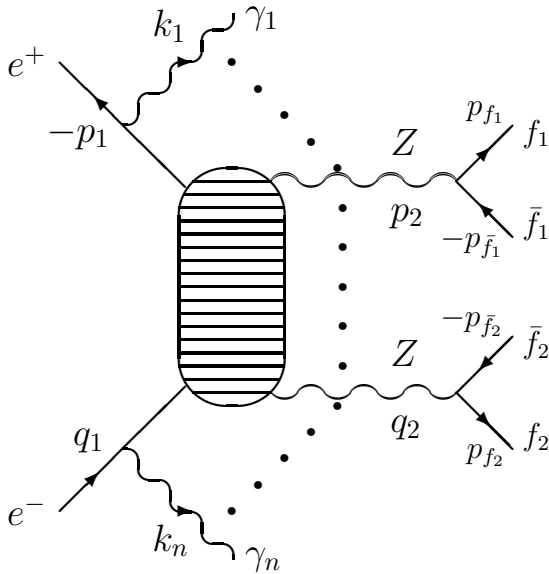

Specifically, in this paper, we extend the result in ref.~\cite{yfsww2}
to the process $e^++e^-\rightarrow Z+Z+n(\gamma)\rightarrow 4 fermions
+n(\gamma)$, as it is illustrated in fig.~\ref{fig:ZZprod}.
We start with the respective master formula.
\par
Referring to our master formula in ref.~\cite{yfsww2}, eq.(9)
in this last reference,
for the process $e^++e^-\rightarrow W^++W^-+n(\gamma)\rightarrow 4 fermions~
+n(\gamma)$ and to the kinematics in fig.~\ref{fig:ZZprod}, we
arrive at the corresponding initial state YFS exponentiated
${\cal O}(\alpha^2)$ leading log (LL) master formula for the
process $e^++e^-\rightarrow Z+Z+n(\gamma)\rightarrow 4 fermions~+
n(\gamma)$ by substituting the respective Born level differential
distribution for the latter process for the former, yielding
\begin{equation}
d\sigma=e^{2\alpha\,Re\,B+2\alpha\,
\tilde B}\sum_{n=0}^\infty{1\over n!}\int\prod_{j=1}^n{d^3k_j\over k_j^0
}\int{d^4y\over(2\pi)^4}\,e^{iy(p_1+q_1-p_2-q_2-\sum_jk_j)+D}\nonumber\cr
\qquad\bar\beta_n(k_1,\dots,k_n){d^3p_{f_1}d^3p_{\bar{f}_1}d^3p_{f_2}
d^3p_{\bar{f}_2}
\over p_{f_1}^0p_{\bar f_1}^0p_{f_2}^0p_{\bar{f}_2}^0},
\label{eq1}
\end{equation}
where we have introduced the same notation as that used in ref.~\cite{yfsww2}
for the YFS infrared functions $B, \tilde B, D$ and for the 
YFS hard photon residuals $\bar\beta_n$. Thus, at the $\bar\beta_0$ level
to which we work 
(i.e. $\bar\beta_1$ and $\bar\beta_2$ are included here in the LL 
approximation only), 
we may identify ${1\over 2}\bar\beta_0$ with 
$d\sigma_{Born}/d\Omega_1d\Omega_2$ where $d\Omega_{1(2)}$ is the
differential decay solid angle of $f_1(f_2)$ in the respective $Z$'s
rest frame. Our respective Born cross section $d\sigma_{Born}$ 
is taken from ref.~\cite{hagi}, where we allow for anomalous couplings
gauge boson  couplings $f_4, f_5$ in the notation of this latter reference.
We have realized the result (\ref{eq1}) using MC methods of two of
us (S.J. and B.F.L.W.~\cite{sjw,yfs2,BHLUMI-89}),
in complete analogy with our work in ref.~\cite{yfsww2}. The resulting
program is called YFSZZ~1.0 and it is available from the authors \cite{yfszz1}.
We will now illustrate
some of its applications. Such multiple photon radiative effects in
the process under study here have not appeared elsewhere.\par

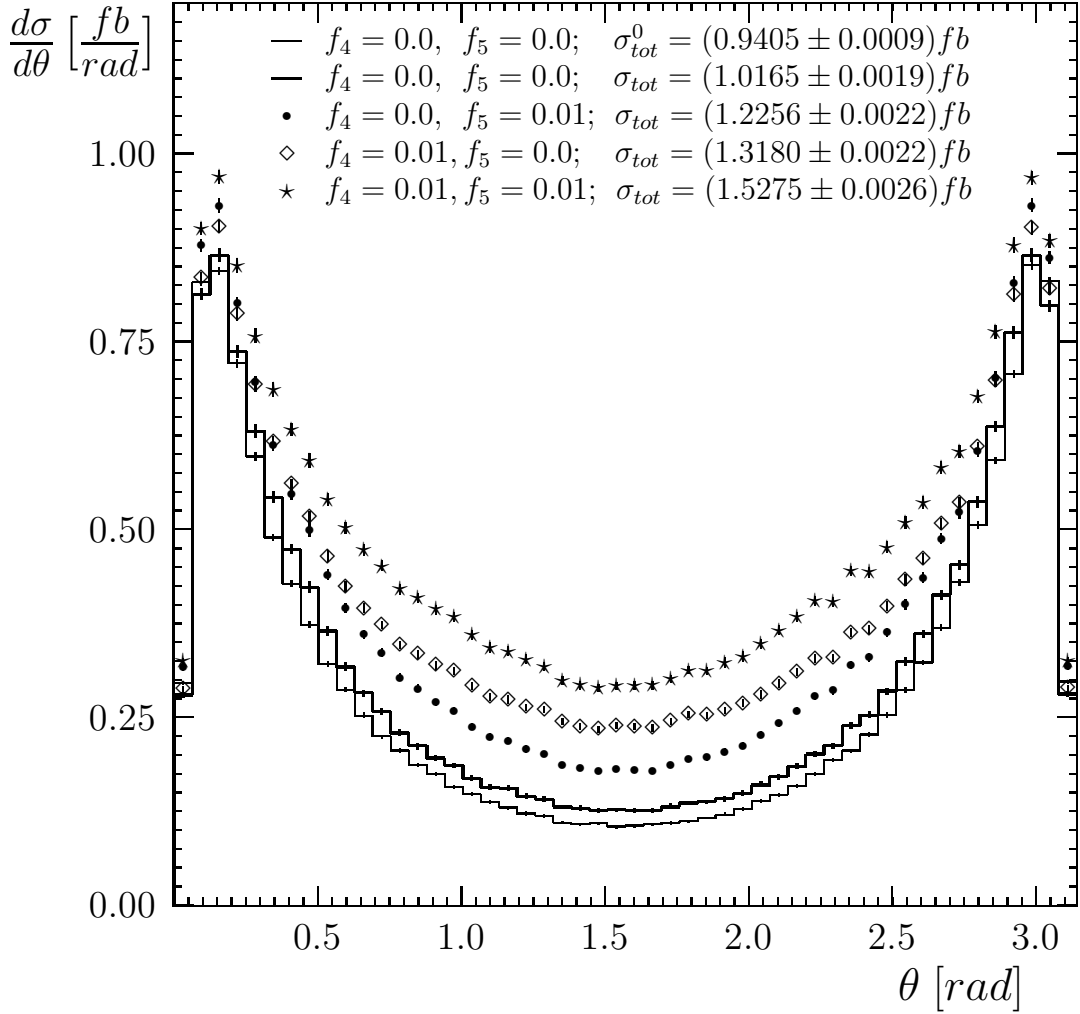
\begin{figure}[!ht]
\centering
\setlength{\unitlength}{0.1mm}
\begin{picture}(1600,1500)
\put(300,250){\begin{picture}( 1200,1200)
\put(0,0){\framebox( 1200,1200){ }}
\multiput(  190.99,0)(  190.99,0){   6}{\line(0,1){25}}
\multiput(     .00,0)(   19.10,0){  63}{\line(0,1){10}}
\multiput(  190.99,1200)(  190.99,0){   6}{\line(0,-1){25}}
\multiput(     .00,1200)(   19.10,0){  63}{\line(0,-1){10}}
\put( 190,-25){\makebox(0,0)[t]{\large $    0.5 $}}
\put( 381,-25){\makebox(0,0)[t]{\large $    1.0 $}}
\put( 572,-25){\makebox(0,0)[t]{\large $    1.5 $}}
\put( 763,-25){\makebox(0,0)[t]{\large $    2.0 $}}
\put( 954,-25){\makebox(0,0)[t]{\large $    2.5 $}}
\put(1145,-25){\makebox(0,0)[t]{\large $    3.0 $}}
\put(1046,-80){\makebox(0,0)[t]{\Large $ \theta\;[rad] $}}
\multiput(0,     .00)(0,  250.00){   5}{\line(1,0){25}}
\multiput(0,   25.00)(0,   25.00){  48}{\line(1,0){10}}
\multiput(1200,     .00)(0,  250.00){   5}{\line(-1,0){25}}
\multiput(1200,   25.00)(0,   25.00){  48}{\line(-1,0){10}}
\put(-25,   0){\makebox(0,0)[r]{\large $    0.00 $}}
\put(-25, 250){\makebox(0,0)[r]{\large $    0.25 $}}
\put(-25, 500){\makebox(0,0)[r]{\large $    0.50 $}}
\put(-25, 750){\makebox(0,0)[r]{\large $    0.75 $}}
\put(-25,1000){\makebox(0,0)[r]{\large $    1.00 $}}
\put(-25,1150){\makebox(0,0)[r]{\LARGE $ \frac{d\sigma}{d\theta}\,
                                         [\frac{fb}{rad}] $}}
\put(150,850){\begin{picture}( 300,900)
 \thinlines
 \put(-20,300){\line(1,0){40} }
 \put(50,300){\makebox(0,0)[l]{$f_4=0.0,\:\: f_5=0.0;\:\:\:\:
                               \sigma_{tot}^0 =(0.9405 \pm 0.0009)fb$ }}
 \thicklines
 \put(-20,250){\line(1,0){40} } 
 \thinlines
 \put(50,250){\makebox(0,0)[l]{$f_4=0.0,\:\: f_5=0.0;\:\:\:\:
                               \sigma_{tot}=(1.0165 \pm 0.0019)fb$ }}
 \put( 5,200){\makebox(0,0){\circle*{ 10}}}                                 
 \put(50,200){\makebox(0,0)[l]{$f_4=0.0,\:\: f_5=0.01;\:\:
                               \sigma_{tot}=(1.2256 \pm 0.0022)fb$ }}  
 \put( 0,150){\makebox(0,0){$\diamond$}}  
 \put(50,150){\makebox(0,0)[l]{$f_4=0.01, f_5=0.0;\:\:\:\:
                               \sigma_{tot}=(1.3180 \pm 0.0022)fb$ }} 
 \put( 0,100){\makebox(0,0){$\star$}}                 
 \put(50,100){\makebox(0,0)[l]{$f_4=0.01, f_5=0.01;\:\:
                              \sigma_{tot}=(1.5275 \pm 0.0026)fb$  }}
  
\end{picture}} 
\end{picture}}
\put(300,250){\begin{picture}( 1200,1200)
\thinlines 
\newcommand{\x}[3]{\put(#1,#2){\line(1,0){#3}}}
\newcommand{\y}[3]{\put(#1,#2){\line(0,1){#3}}}
\newcommand{\z}[3]{\put(#1,#2){\line(0,-1){#3}}}
\newcommand{\e}[3]{\put(#1,#2){\line(0,1){#3}}}
\y{   0}{   0}{ 296}\x{   0}{ 296}{  24}
\e{  12}{  294}{   4}
\y{  24}{ 296}{ 533}\x{  24}{ 829}{  24}
\e{  36}{  824}{   8}
\y{  48}{ 829}{  15}\x{  48}{ 844}{  24}
\e{  60}{  839}{  10}
\z{  72}{ 844}{ 123}\x{  72}{ 721}{  24}
\e{  84}{  716}{  10}
\z{  96}{ 721}{ 124}\x{  96}{ 597}{  24}
\e{ 108}{  592}{  10}
\z{ 120}{ 597}{ 108}\x{ 120}{ 489}{  24}
\e{ 132}{  485}{   8}
\z{ 144}{ 489}{  61}\x{ 144}{ 428}{  24}
\e{ 156}{  424}{   8}
\z{ 168}{ 428}{  55}\x{ 168}{ 373}{  24}
\e{ 180}{  369}{   8}
\z{ 192}{ 373}{  52}\x{ 192}{ 321}{  24}
\e{ 204}{  318}{   6}
\z{ 216}{ 321}{  34}\x{ 216}{ 287}{  24}
\e{ 228}{  284}{   6}
\z{ 240}{ 287}{  35}\x{ 240}{ 252}{  24}
\e{ 252}{  250}{   6}
\z{ 264}{ 252}{  27}\x{ 264}{ 225}{  24}
\e{ 276}{  222}{   4}
\z{ 288}{ 225}{  19}\x{ 288}{ 206}{  24}
\e{ 300}{  204}{   4}
\z{ 312}{ 206}{  19}\x{ 312}{ 187}{  24}
\e{ 324}{  185}{   4}
\z{ 336}{ 187}{  12}\x{ 336}{ 175}{  24}
\e{ 348}{  173}{   4}
\z{ 360}{ 175}{  18}\x{ 360}{ 157}{  24}
\e{ 372}{  155}{   4}
\z{ 384}{ 157}{   9}\x{ 384}{ 148}{  24}
\e{ 396}{  146}{   4}
\z{ 408}{ 148}{  10}\x{ 408}{ 138}{  24}
\e{ 420}{  136}{   4}
\z{ 432}{ 138}{   8}\x{ 432}{ 130}{  24}
\e{ 444}{  128}{   4}
\z{ 456}{ 130}{   8}\x{ 456}{ 122}{  24}
\e{ 468}{  120}{   4}
\z{ 480}{ 122}{   3}\x{ 480}{ 119}{  24}
\e{ 492}{  117}{   4}
\z{ 504}{ 119}{   9}\x{ 504}{ 110}{  24}
\e{ 516}{  109}{   2}
\z{ 528}{ 110}{   2}\x{ 528}{ 108}{  24}
\e{ 540}{  106}{   2}
\y{ 552}{ 108}{   1}\x{ 552}{ 109}{  24}
\e{ 564}{  107}{   2}
\z{ 576}{ 109}{   4}\x{ 576}{ 105}{  24}
\e{ 588}{  103}{   2}
\y{ 600}{ 105}{   1}\x{ 600}{ 106}{  24}
\e{ 612}{  104}{   2}
\y{ 624}{ 106}{   2}\x{ 624}{ 108}{  24}
\e{ 636}{  106}{   2}
\y{ 648}{ 108}{   2}\x{ 648}{ 110}{  24}
\e{ 660}{  108}{   4}
\y{ 672}{ 110}{   2}\x{ 672}{ 112}{  24}
\e{ 684}{  111}{   2}
\y{ 696}{ 112}{   4}\x{ 696}{ 116}{  24}
\e{ 708}{  115}{   2}
\y{ 720}{ 116}{   4}\x{ 720}{ 120}{  24}
\e{ 732}{  119}{   4}
\y{ 744}{ 120}{   8}\x{ 744}{ 128}{  24}
\e{ 756}{  126}{   4}
\y{ 768}{ 128}{  11}\x{ 768}{ 139}{  24}
\e{ 780}{  137}{   4}
\y{ 792}{ 139}{   8}\x{ 792}{ 147}{  24}
\e{ 804}{  145}{   4}
\y{ 816}{ 147}{  12}\x{ 816}{ 159}{  24}
\e{ 828}{  157}{   4}
\y{ 840}{ 159}{  16}\x{ 840}{ 175}{  24}
\e{ 852}{  173}{   4}
\y{ 864}{ 175}{  18}\x{ 864}{ 193}{  24}
\e{ 876}{  191}{   4}
\y{ 888}{ 193}{  13}\x{ 888}{ 206}{  24}
\e{ 900}{  203}{   4}
\y{ 912}{ 206}{  21}\x{ 912}{ 227}{  24}
\e{ 924}{  225}{   4}
\y{ 936}{ 227}{  26}\x{ 936}{ 253}{  24}
\e{ 948}{  250}{   6}
\y{ 960}{ 253}{  33}\x{ 960}{ 286}{  24}
\e{ 972}{  283}{   6}
\y{ 984}{ 286}{  37}\x{ 984}{ 323}{  24}
\e{ 996}{  320}{   6}
\y{1008}{ 323}{  46}\x{1008}{ 369}{  24}
\e{1020}{  366}{   8}
\y{1032}{ 369}{  61}\x{1032}{ 430}{  24}
\e{1044}{  426}{   8}
\y{1056}{ 430}{  76}\x{1056}{ 506}{  24}
\e{1068}{  502}{   8}
\y{1080}{ 506}{  86}\x{1080}{ 592}{  24}
\e{1092}{  588}{   8}
\y{1104}{ 592}{ 115}\x{1104}{ 707}{  24}
\e{1116}{  702}{  10}
\y{1128}{ 707}{ 145}\x{1128}{ 852}{  24}
\e{1140}{  846}{  10}
\z{1152}{ 852}{  21}\x{1152}{ 831}{  24}
\e{1164}{  827}{   8}
\z{1176}{ 831}{ 533}\x{1176}{ 298}{  24}
\e{1188}{  295}{   4}
\end{picture}} 

\put(300,250){\begin{picture}( 1200,1200)
\thicklines 
\newcommand{\x}[3]{\put(#1,#2){\line(1,0){#3}}}
\newcommand{\y}[3]{\put(#1,#2){\line(0,1){#3}}}
\newcommand{\z}[3]{\put(#1,#2){\line(0,-1){#3}}}
\newcommand{\e}[3]{\put(#1,#2){\line(0,1){#3}}}
\y{   0}{   0}{ 280}\x{   0}{ 280}{  24}
\e{  12}{  277}{   6}
\y{  24}{ 280}{ 533}\x{  24}{ 813}{  24}
\e{  36}{  806}{  14}
\y{  48}{ 813}{  52}\x{  48}{ 865}{  24}
\e{  60}{  857}{  16}
\z{  72}{ 865}{ 128}\x{  72}{ 737}{  24}
\e{  84}{  729}{  16}
\z{  96}{ 737}{ 106}\x{  96}{ 631}{  24}
\e{ 108}{  623}{  16}
\z{ 120}{ 631}{  88}\x{ 120}{ 543}{  24}
\e{ 132}{  536}{  14}
\z{ 144}{ 543}{  69}\x{ 144}{ 474}{  24}
\e{ 156}{  468}{  12}
\z{ 168}{ 474}{  51}\x{ 168}{ 423}{  24}
\e{ 180}{  416}{  14}
\z{ 192}{ 423}{  58}\x{ 192}{ 365}{  24}
\e{ 204}{  359}{  12}
\z{ 216}{ 365}{  48}\x{ 216}{ 317}{  24}
\e{ 228}{  313}{  10}
\z{ 240}{ 317}{  34}\x{ 240}{ 283}{  24}
\e{ 252}{  278}{   8}
\z{ 264}{ 283}{  25}\x{ 264}{ 258}{  24}
\e{ 276}{  254}{   8}
\z{ 288}{ 258}{  29}\x{ 288}{ 229}{  24}
\e{ 300}{  226}{   8}
\z{ 312}{ 229}{  17}\x{ 312}{ 212}{  24}
\e{ 324}{  208}{   8}
\z{ 336}{ 212}{  16}\x{ 336}{ 196}{  24}
\e{ 348}{  193}{   6}
\z{ 360}{ 196}{  10}\x{ 360}{ 186}{  24}
\e{ 372}{  183}{   6}
\z{ 384}{ 186}{  17}\x{ 384}{ 169}{  24}
\e{ 396}{  166}{   6}
\z{ 408}{ 169}{  12}\x{ 408}{ 157}{  24}
\e{ 420}{  154}{   6}
\z{ 432}{ 157}{   1}\x{ 432}{ 156}{  24}
\e{ 444}{  153}{   6}
\z{ 456}{ 156}{  11}\x{ 456}{ 145}{  24}
\e{ 468}{  142}{   6}
\z{ 480}{ 145}{   4}\x{ 480}{ 141}{  24}
\e{ 492}{  139}{   6}
\z{ 504}{ 141}{  10}\x{ 504}{ 131}{  24}
\e{ 516}{  129}{   4}
\z{ 528}{ 131}{   2}\x{ 528}{ 129}{  24}
\e{ 540}{  126}{   6}
\z{ 552}{ 129}{   3}\x{ 552}{ 126}{  24}
\e{ 564}{  124}{   4}
\y{ 576}{ 126}{   1}\x{ 576}{ 127}{  24}
\e{ 588}{  125}{   4}
\z{ 600}{ 127}{   1}\x{ 600}{ 126}{  24}
\e{ 612}{  124}{   4}
\y{ 624}{ 126}{   0}\x{ 624}{ 126}{  24}
\e{ 636}{  124}{   4}
\y{ 648}{ 126}{   5}\x{ 648}{ 131}{  24}
\e{ 660}{  129}{   6}
\y{ 672}{ 131}{   5}\x{ 672}{ 136}{  24}
\e{ 684}{  134}{   4}
\y{ 696}{ 136}{   2}\x{ 696}{ 138}{  24}
\e{ 708}{  136}{   4}
\y{ 720}{ 138}{   4}\x{ 720}{ 142}{  24}
\e{ 732}{  139}{   6}
\y{ 744}{ 142}{   7}\x{ 744}{ 149}{  24}
\e{ 756}{  146}{   6}
\y{ 768}{ 149}{  11}\x{ 768}{ 160}{  24}
\e{ 780}{  158}{   6}
\y{ 792}{ 160}{  11}\x{ 792}{ 171}{  24}
\e{ 804}{  168}{   6}
\y{ 816}{ 171}{  14}\x{ 816}{ 185}{  24}
\e{ 828}{  182}{   6}
\y{ 840}{ 185}{  16}\x{ 840}{ 201}{  24}
\e{ 852}{  198}{   6}
\y{ 864}{ 201}{  11}\x{ 864}{ 212}{  24}
\e{ 876}{  209}{   6}
\y{ 888}{ 212}{  27}\x{ 888}{ 239}{  24}
\e{ 900}{  235}{   8}
\y{ 912}{ 239}{  14}\x{ 912}{ 253}{  24}
\e{ 924}{  250}{   8}
\y{ 936}{ 253}{  32}\x{ 936}{ 285}{  24}
\e{ 948}{  280}{   8}
\y{ 960}{ 285}{  40}\x{ 960}{ 325}{  24}
\e{ 972}{  319}{  10}
\y{ 984}{ 325}{  36}\x{ 984}{ 361}{  24}
\e{ 996}{  356}{  10}
\y{1008}{ 361}{  52}\x{1008}{ 413}{  24}
\e{1020}{  407}{  12}
\y{1032}{ 413}{  40}\x{1032}{ 453}{  24}
\e{1044}{  447}{  12}
\y{1056}{ 453}{  84}\x{1056}{ 537}{  24}
\e{1068}{  530}{  14}
\y{1080}{ 537}{ 100}\x{1080}{ 637}{  24}
\e{1092}{  630}{  14}
\y{1104}{ 637}{ 125}\x{1104}{ 762}{  24}
\e{1116}{  754}{  16}
\y{1128}{ 762}{ 103}\x{1128}{ 865}{  24}
\e{1140}{  857}{  16}
\z{1152}{ 865}{  67}\x{1152}{ 798}{  24}
\e{1164}{  790}{  14}
\z{1176}{ 798}{ 517}\x{1176}{ 281}{  24}
\e{1188}{  278}{   6}
\end{picture}} 

\put(300,250){\begin{picture}( 1200,1200)
\newcommand{\R}[2]{\put(#1,#2){\circle*{ 10}}}
\newcommand{\eE}[3]{\put(#1,#2){\line(0,1){#3}}}
\R{  12}{ 317}
\eE{  12}{  313}{   8}
\R{  36}{ 878}
\eE{  36}{  870}{  16}
\R{  60}{ 931}
\eE{  60}{  922}{  18}
\R{  84}{ 801}
\eE{  84}{  793}{  16}
\R{ 108}{ 696}
\eE{ 108}{  688}{  16}
\R{ 132}{ 613}
\eE{ 132}{  605}{  16}
\R{ 156}{ 547}
\eE{ 156}{  540}{  14}
\R{ 180}{ 499}
\eE{ 180}{  491}{  16}
\R{ 204}{ 440}
\eE{ 204}{  433}{  14}
\R{ 228}{ 395}
\eE{ 228}{  389}{  12}
\R{ 252}{ 361}
\eE{ 252}{  355}{  10}
\R{ 276}{ 336}
\eE{ 276}{  331}{  10}
\R{ 300}{ 303}
\eE{ 300}{  299}{  10}
\R{ 324}{ 288}
\eE{ 324}{  284}{  10}
\R{ 348}{ 271}
\eE{ 348}{  266}{   8}
\R{ 372}{ 259}
\eE{ 372}{  255}{   8}
\R{ 396}{ 238}
\eE{ 396}{  234}{   8}
\R{ 420}{ 224}
\eE{ 420}{  220}{   8}
\R{ 444}{ 219}
\eE{ 444}{  215}{   8}
\R{ 468}{ 208}
\eE{ 468}{  204}{   8}
\R{ 492}{ 201}
\eE{ 492}{  198}{   6}
\R{ 516}{ 187}
\eE{ 516}{  184}{   6}
\R{ 540}{ 183}
\eE{ 540}{  179}{   6}
\R{ 564}{ 179}
\eE{ 564}{  176}{   6}
\R{ 588}{ 181}
\eE{ 588}{  178}{   6}
\R{ 612}{ 180}
\eE{ 612}{  177}{   6}
\R{ 636}{ 179}
\eE{ 636}{  176}{   6}
\R{ 660}{ 187}
\eE{ 660}{  184}{   6}
\R{ 684}{ 195}
\eE{ 684}{  191}{   6}
\R{ 708}{ 197}
\eE{ 708}{  193}{   6}
\R{ 732}{ 204}
\eE{ 732}{  200}{   8}
\R{ 756}{ 212}
\eE{ 756}{  209}{   8}
\R{ 780}{ 227}
\eE{ 780}{  223}{   8}
\R{ 804}{ 242}
\eE{ 804}{  238}{   8}
\R{ 828}{ 259}
\eE{ 828}{  255}{   8}
\R{ 852}{ 278}
\eE{ 852}{  274}{   8}
\R{ 876}{ 286}
\eE{ 876}{  281}{   8}
\R{ 900}{ 320}
\eE{ 900}{  315}{  10}
\R{ 924}{ 330}
\eE{ 924}{  325}{  10}
\R{ 948}{ 363}
\eE{ 948}{  358}{  10}
\R{ 972}{ 401}
\eE{ 972}{  394}{  12}
\R{ 996}{ 436}
\eE{ 996}{  430}{  12}
\R{1020}{ 488}
\eE{1020}{  481}{  14}
\R{1044}{ 523}
\eE{1044}{  515}{  14}
\R{1068}{ 604}
\eE{1068}{  597}{  16}
\R{1092}{ 702}
\eE{1092}{  694}{  16}
\R{1116}{ 828}
\eE{1116}{  819}{  18}
\R{1140}{ 931}
\eE{1140}{  923}{  18}
\R{1164}{ 862}
\eE{1164}{  854}{  16}
\R{1188}{ 318}
\eE{1188}{  314}{   8}
\end{picture}} 

\put(300,250){\begin{picture}( 1200,1200)
\newcommand{\R}[2]{\put(#1,#2){\makebox(0,0){$\diamond$}}}
\newcommand{\eE}[3]{\put(#1,#2){\line(0,1){#3}}}
\R{  12}{ 288}
\eE{  12}{  285}{   6}
\R{  36}{ 835}
\eE{  36}{  828}{  14}
\R{  60}{ 903}
\eE{  60}{  895}{  16}
\R{  84}{ 787}
\eE{  84}{  778}{  16}
\R{ 108}{ 692}
\eE{ 108}{  684}{  16}
\R{ 132}{ 616}
\eE{ 132}{  608}{  16}
\R{ 156}{ 560}
\eE{ 156}{  553}{  14}
\R{ 180}{ 517}
\eE{ 180}{  508}{  16}
\R{ 204}{ 464}
\eE{ 204}{  457}{  14}
\R{ 228}{ 424}
\eE{ 228}{  418}{  12}
\R{ 252}{ 395}
\eE{ 252}{  389}{  12}
\R{ 276}{ 373}
\eE{ 276}{  368}{  10}
\R{ 300}{ 346}
\eE{ 300}{  341}{  10}
\R{ 324}{ 334}
\eE{ 324}{  329}{  10}
\R{ 348}{ 319}
\eE{ 348}{  314}{  10}
\R{ 372}{ 312}
\eE{ 372}{  307}{  10}
\R{ 396}{ 292}
\eE{ 396}{  287}{  10}
\R{ 420}{ 277}
\eE{ 420}{  272}{  10}
\R{ 444}{ 274}
\eE{ 444}{  269}{  10}
\R{ 468}{ 264}
\eE{ 468}{  259}{   8}
\R{ 492}{ 260}
\eE{ 492}{  256}{   8}
\R{ 516}{ 244}
\eE{ 516}{  240}{   8}
\R{ 540}{ 238}
\eE{ 540}{  234}{   8}
\R{ 564}{ 235}
\eE{ 564}{  230}{   8}
\R{ 588}{ 239}
\eE{ 588}{  235}{   8}
\R{ 612}{ 238}
\eE{ 612}{  234}{   8}
\R{ 636}{ 236}
\eE{ 636}{  232}{   8}
\R{ 660}{ 245}
\eE{ 660}{  241}{   8}
\R{ 684}{ 255}
\eE{ 684}{  250}{   8}
\R{ 708}{ 253}
\eE{ 708}{  248}{   8}
\R{ 732}{ 260}
\eE{ 732}{  255}{   8}
\R{ 756}{ 268}
\eE{ 756}{  264}{  10}
\R{ 780}{ 280}
\eE{ 780}{  275}{  10}
\R{ 804}{ 295}
\eE{ 804}{  290}{  10}
\R{ 828}{ 310}
\eE{ 828}{  305}{  10}
\R{ 852}{ 328}
\eE{ 852}{  323}{  10}
\R{ 876}{ 329}
\eE{ 876}{  324}{  10}
\R{ 900}{ 363}
\eE{ 900}{  358}{  10}
\R{ 924}{ 367}
\eE{ 924}{  362}{  10}
\R{ 948}{ 397}
\eE{ 948}{  392}{  12}
\R{ 972}{ 433}
\eE{ 972}{  426}{  14}
\R{ 996}{ 461}
\eE{ 996}{  454}{  14}
\R{1020}{ 507}
\eE{1020}{  500}{  14}
\R{1044}{ 535}
\eE{1044}{  528}{  14}
\R{1068}{ 610}
\eE{1068}{  602}{  14}
\R{1092}{ 698}
\eE{1092}{  690}{  16}
\R{1116}{ 812}
\eE{1116}{  803}{  18}
\R{1140}{ 901}
\eE{1140}{  893}{  16}
\R{1164}{ 820}
\eE{1164}{  813}{  14}
\R{1188}{ 289}
\eE{1188}{  285}{   8}
\end{picture}} 

\put(300,250){\begin{picture}( 1200,1200)
\newcommand{\R}[2]{\put(#1,#2){\makebox(0,0){$\star$}}}
\newcommand{\eE}[3]{\put(#1,#2){\line(0,1){#3}}}
\R{  12}{ 325}
\eE{  12}{  321}{   8}
\R{  36}{ 900}
\eE{  36}{  892}{  16}
\R{  60}{ 969}
\eE{  60}{  960}{  18}
\R{  84}{ 851}
\eE{  84}{  842}{  18}
\R{ 108}{ 757}
\eE{ 108}{  749}{  18}
\R{ 132}{ 686}
\eE{ 132}{  677}{  18}
\R{ 156}{ 633}
\eE{ 156}{  625}{  16}
\R{ 180}{ 592}
\eE{ 180}{  583}{  18}
\R{ 204}{ 540}
\eE{ 204}{  532}{  16}
\R{ 228}{ 502}
\eE{ 228}{  495}{  14}
\R{ 252}{ 473}
\eE{ 252}{  466}{  14}
\R{ 276}{ 451}
\eE{ 276}{  445}{  14}
\R{ 300}{ 421}
\eE{ 300}{  414}{  12}
\R{ 324}{ 409}
\eE{ 324}{  403}{  12}
\R{ 348}{ 394}
\eE{ 348}{  387}{  12}
\R{ 372}{ 384}
\eE{ 372}{  378}{  12}
\R{ 396}{ 360}
\eE{ 396}{  354}{  12}
\R{ 420}{ 343}
\eE{ 420}{  337}{  12}
\R{ 444}{ 337}
\eE{ 444}{  332}{  12}
\R{ 468}{ 327}
\eE{ 468}{  322}{  12}
\R{ 492}{ 318}
\eE{ 492}{  313}{  10}
\R{ 516}{ 299}
\eE{ 516}{  294}{  10}
\R{ 540}{ 294}
\eE{ 540}{  288}{  10}
\R{ 564}{ 289}
\eE{ 564}{  284}{  10}
\R{ 588}{ 292}
\eE{ 588}{  287}{  10}
\R{ 612}{ 293}
\eE{ 612}{  288}{  10}
\R{ 636}{ 294}
\eE{ 636}{  288}{  10}
\R{ 660}{ 301}
\eE{ 660}{  296}{  10}
\R{ 684}{ 313}
\eE{ 684}{  308}{  10}
\R{ 708}{ 312}
\eE{ 708}{  307}{  10}
\R{ 732}{ 323}
\eE{ 732}{  317}{  12}
\R{ 756}{ 331}
\eE{ 756}{  325}{  12}
\R{ 780}{ 348}
\eE{ 780}{  342}{  12}
\R{ 804}{ 366}
\eE{ 804}{  360}{  12}
\R{ 828}{ 384}
\eE{ 828}{  378}{  12}
\R{ 852}{ 405}
\eE{ 852}{  399}{  12}
\R{ 876}{ 404}
\eE{ 876}{  397}{  12}
\R{ 900}{ 445}
\eE{ 900}{  439}{  12}
\R{ 924}{ 444}
\eE{ 924}{  437}{  12}
\R{ 948}{ 476}
\eE{ 948}{  470}{  14}
\R{ 972}{ 509}
\eE{ 972}{  501}{  16}
\R{ 996}{ 536}
\eE{ 996}{  528}{  16}
\R{1020}{ 582}
\eE{1020}{  574}{  16}
\R{1044}{ 604}
\eE{1044}{  596}{  16}
\R{1068}{ 677}
\eE{1068}{  669}{  16}
\R{1092}{ 763}
\eE{1092}{  754}{  18}
\R{1116}{ 878}
\eE{1116}{  868}{  20}
\R{1140}{ 968}
\eE{1140}{  959}{  18}
\R{1164}{ 884}
\eE{1164}{  876}{  16}
\R{1188}{ 326}
\eE{1188}{  322}{   8}
\end{picture}} 

\end{picture} 

\caption{ Total cross sections and angular distributions for
the process in \protect fig.~\ref{fig:ZZprod} for 
$f_1=e, f_2=\mu$ at $E_{CM}= 500$ GeV. $\sigma^0_{tot}$
corresponds to the respective Born total cross section.
The light solid curve, dark solid curve, solid dot, open diamond,
and solid star correspond respectively to the differential
distributions the polar angle ($\theta$) of one of the $Z$'s relative to the
incoming $e^-$ direction for example for the Standard Model Born cross section,
for the SM ${\cal O}(\alpha^2)$ LL YFS exponentiated cross section,
for the $f_4=0.0, f_5=0.01$ ${\cal O}(\alpha^2)$ LL YFS 
exponentiated cross section, 
for the $f_4=0.01, f_5=0.0$ ${\cal O}(\alpha^2)$ LL YFS 
exponentiated cross section, 
and for the $f_4=0.01, f_5=0.01$ ${\cal O}(\alpha^2)$ LL YFS 
exponentiated cross section, respectively.}
\label{fig:res}
\end{figure}

Specifically, in fig.~\ref{fig:res}, we show the results for $10^6$
simulated (weighted) events 
for the case that $f_1=e,f_2= \mu$ in fig.~\ref{fig:ZZprod}.
We show this for the NLC type CMS energy $E_{CM}= 500$ GeV. To illustrate
the interplay of the $n(\gamma)$ radiation and the anomalous gauge
boson couplings $f_4,f_5$, we calculate five different scenarios,
$f_i=0$, $i=4,5$ (SM gauge boson couplings) at the Born level;
$f_i=0$, $i=4,5$ (SM gauge boson couplings) at the ${\cal O}(\alpha^2)$ LL 
YFS exponentiated $\bar\beta_0$ level;
$f_4=.01$, $f_5=0$; $f_4=0$,
$f_5=.01$; and, finally, $f_4=f_5=.01$,
where the last three scenarios are all
done at the ${\cal O}(\alpha^2)$ LL YFS exponentiated $\bar\beta_0$ level.
The respective total cross sections,
along with the Born cross section denoted by $\sigma^0_{tot}$,
are given in the top of the figure whereas the differential cross
section in the polar production angle of one of the $Z$'s is plotted
in the figure for each case as indicated. The figure illustrates
that these anomalous couplings give angular distributions with general shapes
similar to that of the SM couplings but that they make a significant
increase in the cross section normalization. In this last regard,
the coupling $f_4$ produces a stronger effect than does 
the same strength of the coupling $f_5$; and, when both are
present with the same strength, the effects essentially add linearly 
insofar as the total cross sections are concerned.
This approximate additive linearity is consistent with 
the fact that $f_4$ is CP violating and
$f_5$ is not. At the level of the differential cross sections,
there is a more detailed interplay between the radiative corrections
and the anomalous couplings so that there are regions near the
forward and backward directions where the effect of $f_5$ exceeds
that of $f_4$ whereas in the complementary region the effect of $f_4$ exceeds
that of $f_5$, for the same strength of the respective couplings.
Evidently, the multiple photon character of the
radiative effects, affecting as it does the normalization 
and the detailed angular profile of the
cross sections in the figure for example, must be taken into account
in any precision analysis of such anomalous gauge boson effects.
YFSZZ~1.0 allows one to do this on an event-by-event basis.\par

In ref.~\cite{dima}, semi-analytical results are presented for
the process in fig.~\ref{fig:ZZprod} which feature the exact
${\cal O}(\alpha)$ ISR correction with soft photon exponentiation. 
Indeed, the authors in ref.~\cite{dima}
have isolated both the complete ${\cal O}(\alpha)$ correction and the
so-called universal part of this correction associated with
structure function evolution of the incoming $e^\pm$ beams.
For the process which we present in fig.~\ref{fig:res}, we may therefore
compare directly the ratio the $\sigma_{tot}/\sigma^0_{tot}$
for the case of SM couplings, $1.081 \pm 0.002$, with the analogous ratio
in ref.~\cite{dima} for the respective universal correction, which
is $1.0798$. The excellent agreement 
between these two numbers gives us an important cross check on our
work.\par

In summary, we have developed and illustrated the first ever
multiple photon, finite $p_T$, 
amplitude based Monte Carlo event generator, YFSZZ~1.0,
for the process $e^+e^- \rightarrow
Z Z + n(\gamma)\rightarrow  f_1\bar f_1 + \bar f_2 f_2 + n(\gamma)$
in which infrared singularities are cancelled to all orders in $\alpha$.
We find that there is an interplay between the $n(\gamma)$ radiation
and the presence of possible anomalous couplings in the gauge
boson sector. We look forward with excitement to further
applications of our calculation in high energy $e^+e^-$
colliding beam devices.\par

\vspace{7mm}
\noindent 
{\large\bf  Acknowledgements}

Two of us (S. J. and B.F.L. W.) acknowledge the
kind hospitality of Prof. G. Veneziano and the CERN Theory 
Division while this work
was completed.


\newpage
\noindent
{\Large\bf Figure Captions}\\
\noindent
Fig.~1.  The process
\protect$e^+e^-\rightarrow ZZ + n(\gamma)
\rightarrow  f_1+\bar{f_1} +\bar{f_2} + f_2 + n(\gamma)$.\par 
\noindent
Fig.~2.  Total cross sections and angular distributions for
the process in \protect fig.~\ref{fig:ZZprod} for 
$f_1=e, f_2=\mu$ at $E_{CM}= 500$ GeV. $\sigma^0_{tot}$
corresponds to the respective Born total cross section.
The light solid curve, dark solid curve, solid dot, open diamond,
and solid star correspond respectively to the differential
distributions the polar angle ($\theta$) of one of the $Z$'s relative to the
incoming $e^-$ direction for example for the Standard Model Born cross section,
for the SM ${\cal O}(\alpha^2)$ LL YFS exponentiated cross section,
for the $f_4=0.0, f_5=0.01$ ${\cal O}(\alpha^2)$ LL YFS 
exponentiated cross section, 
for the $f_4=0.01, f_5=0.0$ ${\cal O}(\alpha^2)$ LL YFS 
exponentiated cross section, 
and for the $f_4=0.01, f_5=0.01$ ${\cal O}(\alpha^2)$ LL YFS 
exponentiated cross section, respectively.\par
\end{document}